\documentclass[twocolumn,trackchanges]{aastex701}

\usepackage{graphicx}
\usepackage{amsmath}
\usepackage{txfonts}
\newcommand{\hrieuv}{\text{HRI}_{\text{EUV}}}
\usepackage{bm}
\usepackage{natbib}
\usepackage{mathtools}
\shorttitle{A universal scaling of QPPs}
\shortauthors{Lim et al.}

\submitjournal{ApJL}

\begin{document}

\title{A universal scaling between damping time and period of quasi-periodic pulsations from solar EUV brightenings to X-ray stellar flares}

\author[orcid=0000-0001-9914-9080,sname='Lim']{Daye Lim}
\altaffiliation{Current address: Department of Mathematics and Statistics, University of Exeter, Exeter, EX4 4QF, UK}
\affiliation{Solar-Terrestrial Centre of Excellence – SIDC, Royal Observatory of Belgium, Ringlaan -3- Av. Circulaire, 1180 Brussels, Belgium}
\affiliation{Centre for mathematical Plasma Astrophysics, Department of Mathematics, KU Leuven, Celestijnenlaan 200B, 3001 Leuven, Belgium}
\email[show]{d.lim2@exeter.ac.uk}

\author[orcid=0000-0001-9628-4113,sname='Van Doorsselaere']{Tom Van Doorsselaere} 
\affiliation{Centre for mathematical Plasma Astrophysics, Department of Mathematics, KU Leuven, Celestijnenlaan 200B, 3001 Leuven, Belgium}
\email{tom.vandoorsselaere@kuleuven.be}

\author[orcid=0000-0001-6423-8286,sname=Nakariakov]{Valery M. Nakariakov}
\affiliation{Centre for Fusion, Space and Astrophysics, Physics Department, University of Warwick, Coventry CV4 7AL, UK}
\affiliation{Engineering Research Institute \lq\lq Ventspils International Radio Astronomy Centre (VIRAC)\rq\rq, Ventspils University of Applied Sciences, Ventspils, LV-3601, Latvia}
\affiliation{Centro de Investigacion en Astronomía, Universidad Bernardo O’Higgins, Avenida Viel 1497, Santiago, Chile}
\email{V.Nakariakov@warwick.ac.uk}

\author[orcid=0000-0002-0735-4501,sname='Krishna Prasad']{S. Krishna Prasad}
\affiliation{Aryabhatta Research Institute of Observational Sciences (ARIES), Manora Peak, Nainital-263001, Uttarakhand, India}
\email{krishna.prasad@aries.res.in}  

\author[orcid=0000-0003-4052-9462,sname='Berghmans']{David Berghmans}
\affiliation{Solar-Terrestrial Centre of Excellence – SIDC, Royal Observatory of Belgium, Ringlaan -3- Av. Circulaire, 1180 Brussels, Belgium}
\email{david.berghmans@oma.be}  

\author[orcid=0000-0002-6835-2390,sname='Hayes']{Laura A. Hayes}
\affiliation{Astronomy \& Astrophysics Section, School of Cosmic Physics, Dublin Institute for Advanced Studies, DIAS Dunsink Observatory, Dublin, D15 XR2R, Ireland}
\email{laura.hayes@dias.ie}  

\author[orcid=0000-0003-2161-9606,sname='Cho']{Kyung-Suk Cho}
\affiliation{Space Science Division, Korea Astronomy and Space Science Institute, Daejeon 34055, Republic of Korea}
\affiliation{Department of Astronomy and Space Science, University of Science and Technology, Daejeon 34113, Republic of Korea}
\email{kscho@kasi.re.kr}  

\author[orcid=0000-0002-5004-7734,sname='Kim']{Sujin Kim}
\affiliation{Space Science Division, Korea Astronomy and Space Science Institute, Daejeon 34055, Republic of Korea}
\email{sjkim@kasi.re.kr}  

\correspondingauthor{Daye Lim}

\begin{abstract}

Recent high spatial and temporal resolution extreme-ultraviolet (EUV) imaging observations have revealed that quasi-periodic pulsations (QPPs), a ubiquitous signature of impulsive energy release in solar and stellar flares, are also present in much smaller-scale coronal events known as EUV brightenings. Whether QPPs observed across such disparate spatial and energetic scales share a common physical origin remains an open question. Here we analyse 2,146 EUV brightenings observed with Solar Orbiter/EUI and 300 EUV solar flares observed with SDO/AIA, identifying 185 brightenings and 89 flares exhibiting statistically significant damped QPPs. We show that the relationship between damping time and oscillation period follows a common power-law scaling for EUV brightenings and EUV solar flares, consistent with previously reported X-ray QPPs spanning both solar and stellar flares. The persistence of this scaling over a wide range of energies and scales suggests that QPPs are governed by a common underlying physical mechanism. 

\end{abstract}


\section{Introduction} 

Since the launch of the Solar and Heliospheric Observatory and its Extreme ultraviolet Imaging Telescope \citep{1995SoPh..162..291D} three decades ago, imaging of the solar corona has revealed that it is permeated with small-scale events exhibiting sudden, transient intensity enhancements, commonly referred to as extreme-ultraviolet (EUV) brightenings \citep{1998A&A...336.1039B, 1998SoPh..182..349B, 1999SoPh..186..207B}. With the high spatial and temporal resolution of the Solar Orbiter/Extreme Ultraviolet Imager (EUI; \citealt{2020A&A...642A...8R}), it has become clear that the corona hosts an enormous number of extremely small (down to 100~km; \citealt{2025A&A...699A.138N}) and rapid (as short as 1~s; \citealt{2025A&A...704A..58L}) brightenings. Because EUV brightenings have been proposed as potential signatures of nanoflares that may play a crucial role in coronal heating, it is important to determine whether they originate from mechanisms analogous to those driving more powerful flares. However, this question remains unresolved, as some EUV brightenings show indications that could be associated with magnetic reconnection driven processes \citep{2021A&A...656L...7C, 2021ApJ...921L..20P, 2022A&A...660A.143K, 2022SoPh..297..141B}, whereas others point towards alternative physical processes \citep{2024A&A...692A.236N, 2024ApJ...969L..34K}.

One of the most distinctive observational signatures in solar and stellar flares is the presence of quasi-periodic pulsations (QPPs). These manifest as fluctuations in the flare emission that contain an oscillatory component. QPPs have been detected across an exceptionally wide range of wavelengths, from radio to gamma-rays \citep[see, e.g.,][for detailed reviews]{2010PPCF...52l4009N, 2016SoPh..291.3143V, 2019PPCF...61a4024N, 2021SSRv..217...66Z}, and in some cases exhibit consistent oscillation periods simultaneously across multiple wavelengths \citep{2015ApJ...807...72L, 2021ApJ...910..123C, 2024ApJ...977..207F, 2025ApJS..281...46L}. In solar flares, reported QPP periods span from fractions of a second to several tens of minutes, while in stellar flares they can extend to several hours \citep{2019ApJ...884..160V}. Moreover, QPPs are not confined to the impulsive phase of flares, but are also frequently observed during the decay phase \citep{2016ApJ...827L..30H, 2023MNRAS.523.3689M, 2025ApJ...993...99L}. The term QPP therefore encompasses a range of phenomenologically distinct behaviours. A variety of physical processes have been proposed to account for QPPs, and broadly, they can be classified into three categories: magnetohydrodynamic (MHD) wave phenomena, spontaneous repetitive magnetic reconnection (often termed magnetic dripping), and reconnection that is periodically modulated by external MHD waves \citep{2018SSRv..214...45M, 2020STP.....6a...3K}. Statistical analyses indicate that QPPs occur frequently and may, in fact, represent an inherent property of flares \citep[e.g.,][]{2015SoPh..290.3625S, 2018SoPh..293...61D, 2020ApJ...895...50H}.

Recent studies \citep{2025A&A...698A..65L, 2025A&A...704A..58L} have begun to investigate whether QPP signatures are also present in EUV brightenings, and emerging evidence now suggests that these small-scale events can indeed exhibit quasi-periodic behaviour. In line with findings for QPPs in solar \citep{2019A&A...624A..65P, 2020ApJ...895...50H} and stellar flares \citep{2016MNRAS.459.3659P, 2025A&A...700A.178J}, no correlation has been identified between the QPP period and the peak brightness of EUV brightenings, and comparable trends are observed between the QPP period and event lifetime. These similarities support the interpretation that EUV brightenings may represent small-scale manifestations of flares, motivating further investigation into whether comparable scaling relations apply.

\citet{2016ApJ...830..110C} showed that solar and stellar flares in X-ray emission exhibit nearly identical scaling relations between the damping time and oscillation period for exponentially damped QPPs observed in the decay phase, suggesting that these pulsations may arise from MHD slow or kink modes. This naturally raises the question of whether similar behaviour also extends to much smaller events such as EUV brightenings. In this Letter, we demonstrate universal scaling laws for decay phase damped QPPs detected in EUV brightenings and solar flares in EUV emission, suggesting a shared physical origin.

\section{Data and methods}

\begin{figure}
   \centering
   \includegraphics[width=\hsize]{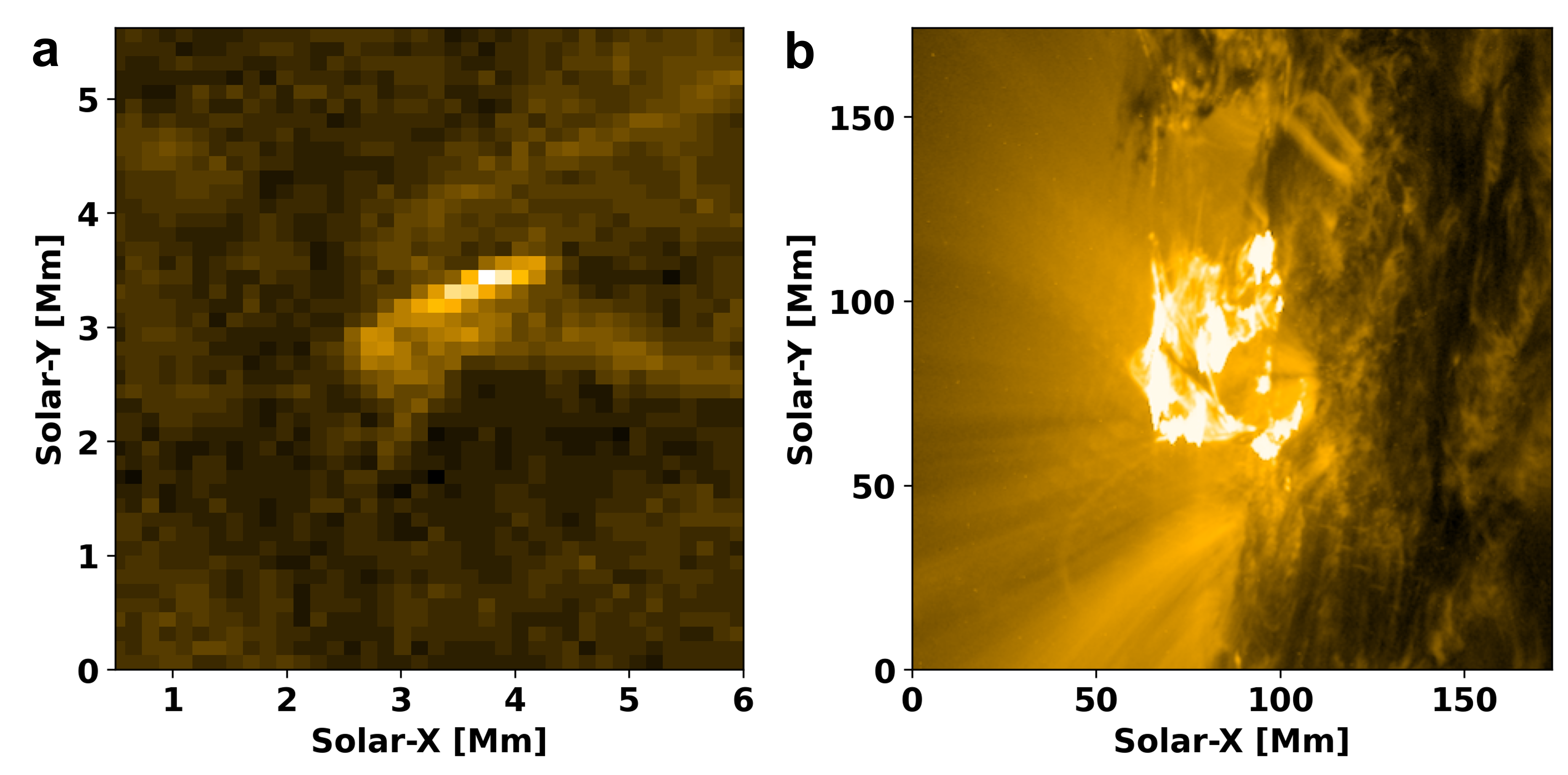}
      \caption{Representative examples of analysed events. (a) An EUV brightening observed on 19 March 2025 at 11:26:38 UT with the Solar Orbiter/EUI $\hrieuv$. (b) An X-class solar flare observed on 3 January 2025 at 22:37:33 UT in the SDO/AIA 171 \AA.}
         \label{fig:events}
   \end{figure}

Between 11:00:01 and 11:28:01 UT on 2025 March 19, EUV brightenings in quiet Sun regions were observed with the Solar Orbiter/EUI High Resolution Imager at 174~\AA\ ($\hrieuv$). The high spatial and temporal resolution of the observations, corresponding to a pixel plate scale of approximately 140~km and a cadence of 1~s, enabled small-scale events and short-lived features to be resolved. These brightenings were identified using an automated detection algorithm \citep{2021A&A...656L...4B, 2025A&A...699A.138N} that isolates transient intensity enhancements exceeding the instrumental noise level \citep{2025A&A...704A..58L}. In that study, 2,146 events were reported to exhibit QPP signatures. For each event, a normalised light curve was constructed by integrating the EUV intensity over the brightening region throughout its catalogue-defined lifetime. The present work re-analyses this QPP subset, focusing on their decay phase light curves to determine characteristic oscillation periods and damping times (Fig.~\ref{fig:events}a). 

A direct comparison with X-ray flare scaling laws \citep{2016ApJ...830..110C} is not straightforward because of the substantially different temperature regimes. To provide an EUV-based reference, we therefore analyse solar flares observed in EUV emission. For this purpose, we use the list\footnote{\url{https://www.lmsal.com/solarsoft/latest_events_archive.html}} of solar flares recorded by the Geostationary Operational Environmental Satellite (GOES) X-ray between 2020 January 1 and 2026 June 7, with flare locations provided in the SolarSoft Latest Events Archive from the Lockheed Martin Solar and Astrophysics Laboratory. To obtain a balanced flare sample while avoiding the statistics being dominated by the far more numerous C-class events, we selected equal numbers of C-, M-, and X-class flares. All available X-class flares during this period were included, while 100 C-class and 100 M-class flares were randomly selected, resulting in a total sample of 300 solar flares.

\begin{figure*}
   \centering
   \includegraphics[width=\hsize]{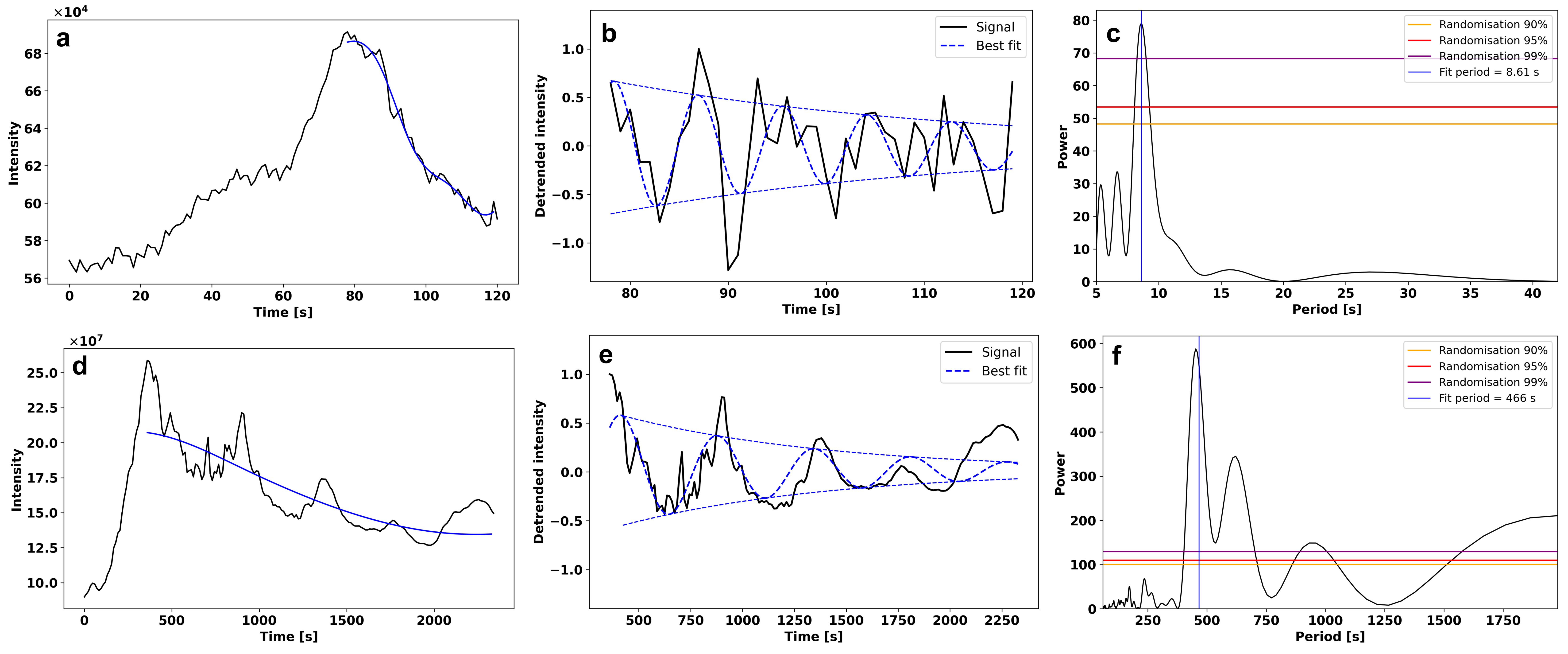}
      \caption{Time-series analysis of the events illustrated in Fig.~\ref{fig:events}. Panels (a--c) show the EUV brightening, and panels (d--f) show the solar flare. Panels (a,d): Example EUV light curves (black), with background trends derived using empirical mode decomposition (blue solid curves). Panels (b,e): Detrended signals composed of intrinsic mode functions, together with the best-fitting damped harmonic oscillation models (blue dashed curves). Panels (c,f): Power spectra of the detrended signals. The vertical blue line marks the fitted period, while the purple, red, and yellow curves denote the 99\%, 95\%, and 90\% confidence levels, respectively.}
         \label{fig:analysis}
   \end{figure*}

For each event, 171~\AA\ data from the Solar Dynamics Observatory/Atmospheric Imaging Assembly (SDO/AIA; \citealt{2012SoPh..275...17L}) at its nominal 12~s cadence were downloaded using SunPy \citep{sunpy_community2020} over the flare interval defined in the GOES event list. The data were processed to Level~1.5 using the \textit{aia\_prep.pro} routine in SolarSoftWare \citep{1998SoPh..182..497F}. We then construct a normalised light curve for each flare by integrating the EUV intensity over the region encompassing the flare emission. Examples of an EUV brightening and an EUV solar flare are shown in Fig.~\ref{fig:events}. Their corresponding light curves are presented in the left panels of Fig.~\ref{fig:analysis}.

To investigate QPPs during the decay phase, we define the decay phase of each EUV brightening and EUV solar flare as the interval following the time of peak intensity in the EUV light curve, and exclude events for which this decay interval is shorter than five time frames. Following the detection procedure described in \citet{2016ApJ...830..110C}, thereby enabling a direct comparison with their X-ray results, we apply the empirical mode decomposition (EMD; \citealt{1998RSPSA.454..903H}) method to each decay-phase light curve, a technique known to be well suited for identifying non-stationary QPPs \citep{2015A&A...574A..53K, 2019PPCF...61a4024N, 2022SSRv..218....9A}. The intrinsic mode function (IMF) with the longest characteristic timescale is taken to represent the slowly varying background trend, following common practice in EMD analyses, and is removed from the signal (Fig.~\ref{fig:analysis}a,d). Here, the background represents the large-scale intensity evolution of the event associated with the impulsive brightening and subsequent decay, rather than the oscillatory component itself. Removing this slowly varying trend allows the analysis to focus on the shorter-timescale fluctuations about the background level (Fig.~\ref{fig:analysis}b,e), which are subsequently examined for damped oscillatory behaviour. Although alternative definitions of the background are possible, adopting the longest-period IMF provides a consistent and objective separation of the slowly varying and oscillatory components across all events.

The sum of the residual IMFs is then fitted with a damped harmonic function of the form, 
\begin{equation}
I=A\text{exp}\left(-\frac{t-t_{0}}{\tau}\right)\text{sin}\left(\frac{t-t_{0}}{P}-B\right), 
\end{equation}
where $A$, $\tau$, $t_{0}$, $P$, and $B$ are the amplitude, damping time, starting time, period, and phase of the oscillation, respectively. The corresponding fits are shown in Fig.~\ref{fig:analysis}b,e. Finally, a Fourier transform is applied to the summed residual IMFs to obtain the power spectrum. Figs.~\ref{fig:analysis}c and \ref{fig:analysis}f illustrate an example spectrum together with its 99\%, 95\%, and 90\% significance levels. The significance levels are calculated using the Fisher randomisation method \citep{1985AJ.....90.2317L}, with confidence intervals at each frequency determined from the cumulative distribution of 10,000 noise realisations. 
Only oscillations satisfying the following criteria were retained for subsequent analysis. First, the spectral power at the fitted period was required to exceed the 90\% significance threshold. Second, the uncertainties in both $P$ and $\tau$ had to be smaller than their corresponding fitted values. Third, the fitted period was required to exceed five times the observational cadence, a more stringent constraint than the standard Nyquist criterion. Fourth, only oscillations comprising at least one complete cycle within the decay phase were retained, quantified by requiring the total decay-phase duration to be greater than or equal to the fitted period. Finally, consistency between time-domain and frequency-domain estimates was imposed by requiring that the period obtained from the damped sine fit differ by less than 20\% from the period corresponding to the maximum power in the Fourier spectrum.

To quantify the relationship between the QPP period $P$ and the damping time $\tau$, we analysed the scaling between these quantities in logarithmic space. For each event, the measured values of $P$ and $\tau$ were transformed to $\log P$ and $\log \tau$, such that a power-law relation of the form $\tau \propto P^{\alpha}$ corresponds to a linear relation in log-log space. As a non-parametric assessment of the strength of the association between $P$ and $\tau$, we computed both the Spearman rank correlation coefficient ($\rho_{S}$). This provides a model-independent measure of monotonic correlation and is less sensitive to outliers and deviations from strict linearity. To infer the scaling exponent and its uncertainty, we performed a Bayesian linear regression \citep{pymc2023} in log-log space. The likelihood was defined such that both measurement uncertainties in $P$ and $\tau$, and an additional intrinsic scatter term were explicitly accounted for. For the published X-ray solar and stellar flare sample, uncertainty estimates for the periods and damping times are not reported \citep{2016ApJ...830..110C}. Therefore, for these events, no measurement uncertainties were included in the regression. The regression model can be written as
\begin{equation}
\log \tau_i \sim \mathcal{N}\left(\log C + \alpha \log P_i,\,
\sigma_{\mathrm{int}}^2 + \sigma_{\log \tau,i}^2 + \alpha^2 \sigma_{\log P,i}^2 \right),
\end{equation}
where $\alpha$ is the power-law index, $C$ is the normalisation, $\sigma_{\mathrm{int}}$ represents the intrinsic scatter, and $\sigma_{\log P,i}$ and $\sigma_{\log \tau,i}$ denote the propagated measurement uncertainties in logarithmic space.
Posterior distributions were sampled using Markov chain Monte Carlo methods, and parameter estimates were derived from the marginal posterior distributions. This approach enables a robust quantification of the scaling relation while accounting for heteroscedastic uncertainties and physical dispersion beyond measurement noise.

To assess whether the two different data sets (between EUV brightenings and EUV flares in Fig.~\ref{fig:scatterplot}c and between EUV events and X-ray events in Fig.~\ref{fig:scatterplot}d) follow a common scaling relation or require distinct power-law indices, we compared two competing models. In the first model, all events were assumed to share a common scaling exponent $\alpha$. In the second model, separate scaling exponents were permitted for the two populations. Model performance was evaluated using Pareto-smoothed importance sampling leave-one-out cross-validation \citep{Vehtari17}, which provides an estimate of the expected log predictive density while penalising unnecessary model complexity. A preference for the shared-slope model indicates that a single scaling relation adequately describes both data sets, whereas a significant improvement in predictive performance for the group-dependent model would suggest statistically meaningful differences between the populations.

\section{Results}

During the decay phase following the peak intensity, a subset of EUV brightenings and EUV flares exhibits single period QPPs with exponentially decaying amplitudes. Figs.~\ref{fig:analysis}a--c show a representative EUV brightening, while Figs.~\ref{fig:analysis}d--f present an example X-class flare. In both cases, the decay phase light curves display oscillatory patterns superimposed on a smoothly varying background (Figs.~\ref{fig:analysis}a,d).
After removal of the background trend, the residual signals reveal coherent oscillations that are well described by a damped sinusoidal form (Figs.~\ref{fig:analysis}b,e). These oscillations are characterised by a single dominant period and an exponential decay of the amplitude, indicative of damped QPPs during the decay phase. The corresponding power spectra exhibit pronounced peaks at the fitted periods that exceed the 95\% significance level (Figs.~\ref{fig:analysis}c,f).
Among the analysed events, 389 EUV brightenings, 33 C-class, 32 M-class, and 49 X-class flares exhibit statistically significant periods during the decay phase. Of these, 185 EUV brightenings, 25 C-class, 25 M-class, and 39 X-class flares additionally display statistically significant damping times. 
These results demonstrate that such single period, exponentially damped QPPs are present during the decay phase of small-scale EUV brightenings and EUV solar flares, consistent with previous reports in X-ray solar and stellar flares \citep{2016ApJ...830..110C}.

\begin{figure}
   \centering
   \includegraphics[width=\hsize]{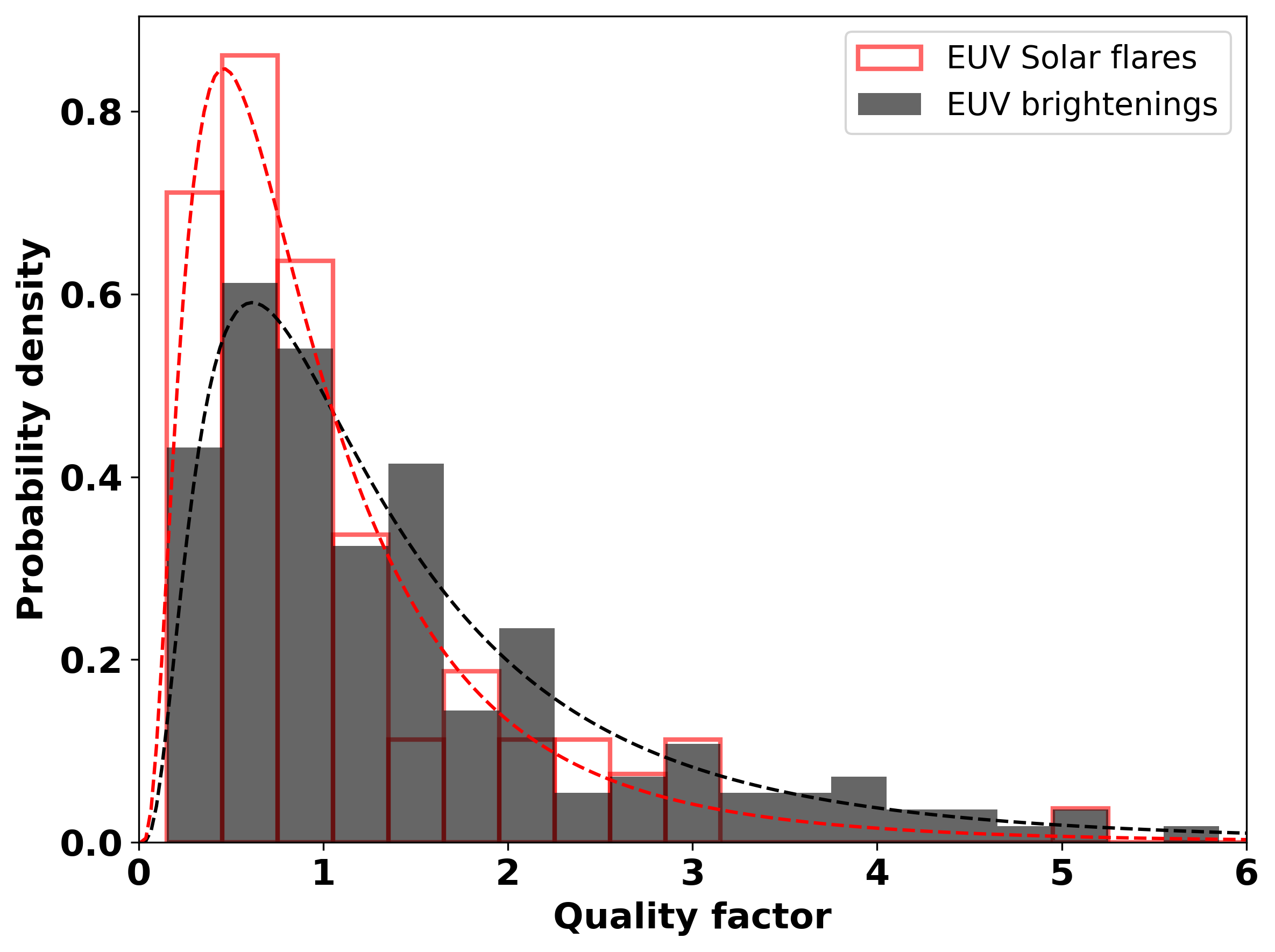}
      \caption{Distributions of QPP quality factors. Probability density distributions of the quality factors for QPPs detected in EUV brightenings (grey) and EUV solar flares (red). The dashed curves indicate the corresponding log-normal distributions, with modal values of 0.5 for solar flares and 0.6 for EUV brightenings.}
         \label{fig:qf}
   \end{figure}

Fig.~\ref{fig:qf} shows the statistical distributions of quality factors, defined as the ratio of the damping time to the oscillation period, for QPPs detected in EUV brightenings and EUV solar flares. The distributions are represented as histograms with a bin width of 0.3. Both distributions are reasonably described by log-normal functions and exhibit similar modal values of 0.6 for EUV brightenings and 0.5 for EUV solar flares. The corresponding mean quality factors are $1.6\pm0.1$ and $1.1\pm0.1$, with standard deviations of 1.7 and 1.3, respectively. A two-sample Kolmogorov--Smirnov test yields a p-value of 0.002, suggesting that the two distributions are statistically distinguishable.

\begin{figure*}
   \centering
   \includegraphics[width=\hsize]{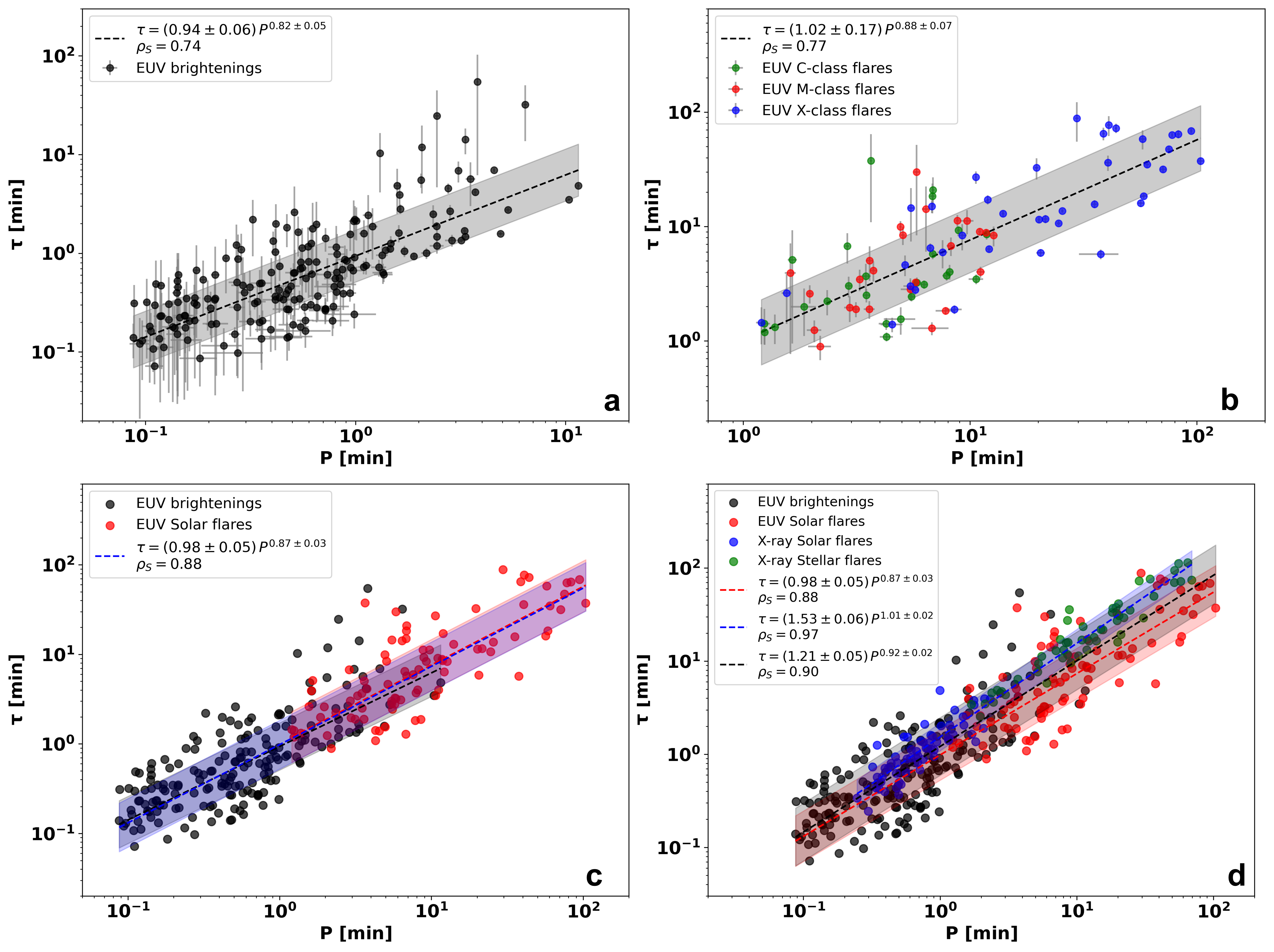}
      \caption{Scaling relation between damping time ($\tau$) and oscillation period ($P$). (a) EUV brightenings, (b) EUV solar flares, (c) the combined EUV events, and (d) the combined EUV data together with previously reported X-ray solar and stellar flares \citep{2016ApJ...830..110C}. Coloured symbols denote different event classes as indicated in the legends. Dashed lines show the median posterior power-law relation inferred from Bayesian linear regression in log-log space. In panel (c), the black and red dashed lines correspond to fits obtained for EUV brightenings and EUV solar flares separately, while the blue dashed line shows the result of a single fit to the combined EUV data set. In panel (d), the red and blue dashed lines represent fits to the EUV and X-ray event samples, respectively, and the black dashed line shows the fit obtained by combining all events. Shaded regions represent the intrinsic scatter associated with each regression. Spearman rank correlation coefficients ($\rho_{S}$) are reported in each panel.}
         \label{fig:scatterplot}
   \end{figure*}

The relationship between the damping time ($\tau$) and the oscillation period ($P$) exhibits a clear monotonic trend across all analysed data sets. This is reflected by statistically significant Spearman rank correlation coefficients ($\rho_{S}$) for EUV brightenings, EUV solar flares, and their combined sample (Fig.~\ref{fig:scatterplot}a--c), indicating that the observed scaling does not rely on a specific functional assumption.

We quantified the $\tau$--$P$ relation using Bayesian linear regression in logarithmic space, explicitly accounting for measurement uncertainties and intrinsic scatter. For EUV brightenings and EUV solar flares analysed separately, the inferred power-law indices are consistent within uncertainties, albeit with substantial intrinsic dispersion. When both EUV populations are combined, the data are well described by a single scaling relation with a power-law index $\alpha \approx 0.87$ (Fig.~\ref{fig:scatterplot}c). Allowing for separate slopes for EUV brightenings and EUV solar flares does not lead to a statistically meaningful improvement in predictive performance, indicating that a shared scaling provides an adequate description of the EUV data.

The EUV results were further compared with previously reported X-ray observations of solar and stellar flares \citep{2016ApJ...830..110C}. When EUV and X-ray events are combined, the full ensemble forms a continuous distribution in the $\tau$--$P$ plane (Fig.~\ref{fig:scatterplot}d). Bayesian model comparison strongly favours a single power-law scaling ($\alpha \approx 0.92$) over models with separate slopes for the EUV and X-ray populations. While the posterior distribution of the slope difference remains broad, reflecting the intrinsic variability of the data, the inclusion of distinct exponents is not statistically required to explain the observations.

Together, these results demonstrate that decay phase damped QPPs observed in EUV brightenings, EUV solar flares, and X-ray solar and stellar flares are compatible with a common scaling between damping time and oscillation period, with intrinsic scatter capturing event-to-event physical variability. The preference for a unified scaling relation across multiple emission regimes supports the interpretation of QPP decay as governed by a broadly universal damping process.

\section{Discussion}

\citet{2016ApJ...830..110C} found that the quality factor distributions of X-ray solar and stellar flares are statistically indistinguishable. Although the quality factor distributions of EUV brightenings and EUV solar flares exhibit similar modal values, the Kolmogorov--Smirnov test provides evidence that the two distributions differ. Inspection of the distributions indicates that this distinction is primarily associated with a more extended high quality factor tail in the EUV brightening population, rather than a systematic shift of the distribution peak. One possible explanation is that a single EUV brightening identified by the automated detection algorithm may occasionally encompass multiple closely spaced flare-like energy-release episodes rather than a single impulsive event. In such cases, additional energy release during the decay phase could repeatedly re-excite the oscillation \citep{2021SSRv..217...73N} and prolong the apparent damping time, contributing to the extended high quality factor tail observed for EUV brightenings. Establishing whether this interpretation is correct is beyond the scope of the present study.

The consistent power-law scaling of damping time with oscillation period across EUV and X-ray observations suggests that the damping of decay phase QPPs may be governed by a common physical process operating over a wide range of energies. Solar flares and other impulsive energy-release events are known to excite a variety of damped wave modes in coronal loops \citep{2021SSRv..217...34W, 2021SSRv..217...73N}. Statistical studies of decaying kink oscillations observed with AIA 171~\AA\ report quality factors in the range 0.5--6.5, with most values exceeding unity, and a power-law index of about 0.6 between the damping time and oscillation period \citep{2019ApJS..241...31N}. Analytical calculations of nonlinear damping in impulsively excited standing kink waves further predict a scaling index close to unity \citep{2021ApJ...910...58V}.  
In this context, the power-law exponents obtained here for decay phase damped QPPs in EUV brightenings ($\sim$0.82) and EUV solar flares ($\sim$0.88) fall between the observational and theoretical expectations for standing kink waves. This consistency suggests that kink-mode dynamics may contribute to the observed damping behaviour.

While kink oscillations provide one plausible interpretation, another class of damped waves commonly observed in coronal loops is standing slow magnetoacoustic oscillations, which are typically detected in hot loops triggered by impulsive events such as flares. A meta-analysis of such oscillations observed over a broad temperature range (0.6-14~MK) reported quality factors of 0.2-4, with roughly half of the values below unity, and a damping time--period scaling index of $\sim$0.9 \citep{2019ApJ...874L...1N}. These statistical properties are closely comparable to those derived here for decay phase damped QPPs in EUV brightenings and EUV solar flares, suggesting that the observed QPPs and their damping may be associated with slow mode oscillations. One of the mechanisms capable of damping slow oscillations is thermal conduction \citep{2002ApJ...580L..85O}. In the strong conduction regime, which may be appropriate for the X-ray flare events, the damping due to thermal conduction is expected to be weaker for slow magnetoacoustic oscillations \citep{2003A&A...408..755D}, resulting in a comparatively large quality factor. The mean quality factor obtained for EUV solar flares in our sample is $1.1$, whereas the corresponding value for solar flares observed in X-rays is $1.7$ \citep{2016ApJ...830..110C}. Assuming comparable loop sizes and densities for the EUV and X-ray flare samples, the higher plasma temperatures diagnosed in X-rays are therefore consistent with the larger quality factor values and comparatively slower damping, in line with conduction-dominated behaviour. Although the quality factor distributions differ between EUV brightenings and EUV flares, this difference does not appear to alter the common period--damping time scaling. Thus, the scaling laws observed across EUV brightenings, solar flares, and stellar flares support an interpretation that their decay phase QPPs and damping processes are related and may reflect slow magnetoacoustic oscillations in flaring or neighbouring coronal loops.

The apparent universality of the period--damping time scaling further supports the interpretation that small-scale EUV brightenings may represent elementary flare events, with potential implications for coronal heating. Moreover, the similarity between the observed scaling relations and those reported for standing kink and slow magnetoacoustic oscillations highlights the diagnostic potential of QPPs as a powerful tool for coronal seismology. Unlike SDO/AIA, the Solar Orbiter/EUI provides high spatial and temporal resolution observations in a single EUV passband, which makes it challenging to directly infer the plasma temperature of individual EUV brightenings \citep{2023A&A...671A..64D, 2025A&A...704A..58L}. In such cases, if the observed decay phase damped QPPs are associated with standing slow oscillations, the measured oscillation period and the characteristic length scale of the brightening can be used to estimate the plasma temperature of the event. 
Similarly, stellar flares lack spatially resolved imaging and therefore do not permit direct measurements of loop parameters. For these unresolved events, the combination of the observed stellar flare temperature and the QPP period can instead be used to infer the characteristic length scale of the flaring loops \citep{2021SoPh..296..162R, 2022ApJ...931...63L}. A more precise application of coronal seismology, however, requires robust identification of the underlying wave mode. Future studies incorporating multi-wavelength observations will be essential for distinguishing between competing MHD interpretations and for establishing more definitive seismological diagnostics across different flare regimes.

%% Please use the acknowledgment and contribution environments. This will 
%% be anonomyized when the "anonymous" style option is used. 
\begin{acknowledgments}
Solar Orbiter is a space mission of international collaboration between ESA and NASA, operated by ESA. The EUI instrument was built by CSL, IAS, MPS, MSSL/UCL, PMOD/WRC, ROB, LCF/IO with funding from the Belgian Federal Science Policy Office (BELSPO/PRODEX PEA C4000134088, 4000112292 and 4000106864); the Centre National d’Etudes Spatiales (CNES); the UK Space Agency (UKSA); the Bundesministerium für Wirtschaft und Energie (BMWi) through the Deutsches Zentrum für Luft- und Raumfahrt (DLR); and the Swiss Space Office (SSO). We thank the numerous team members who have contributed to the success of the SDO mission. DL thanks the Belgian Federal Science Policy Office (BELSPO) for the provision of financial support in the framework of the PRODEX Programme of the European Space Agency (ESA) under contract number 4000143743. TVD was supported by the C1 grant TRACEspace of Internal Funds KU Leuven and a Senior Research Project (G088021N) of the FWO Vlaanderen. Furthermore, TVD received financial support from the Flemish Government under the long-term structural Methusalem funding program, project SOUL: Stellar evolution in full glory, grant METH/24/012 at KU Leuven. The paper is also part of the DynaSun project and has thus received funding under the Horizon Europe programme of the European Union under grant agreement (no. 101131534). Views and opinions expressed are however those of the author(s) only and do not necessarily reflect those of the European Union and therefore the European Union cannot be held responsible for them. VMN acknowledges funding from UK Research and Innovation under the UK government the Horizon Europe funding guarantee EP/Y037456/1 and ERC grant 101201424 (ACDCSUN). SKP is grateful to SERB/ANRF for a startup research grant (No. SRG/2023/002623) and ISRO for a RESPOND grant (No. ISRO/RES/2/453/25-26). KSC acknowledges funding from the Korea Astronomy and Space Science Institute (KASI) under the R\&D program of the Korean government (MSIT: KASI 2026185301). LAH is supported by a Royal Society-Research Ireland University Research Fellowship (URF$\backslash$R1$\backslash$241775).
\end{acknowledgments}

\facilities{SolO, SDO, GOES}

\software{SolarSoftWare \citep{1998SoPh..182..497F},  
          SunPy \citep{sunpy_community2020}, 
          astropy \citep{2013A&A...558A..33A,2018AJ....156..123A,2022ApJ...935..167A},
          SciPy \citep{2020SciPy-NMeth}
          }

%% For this sample we use BibTeX plus aasjournalv7.bst to generate the

\bibliography{Lim_bib}{}
\bibliographystyle{aasjournalv7}



\end{document}